\documentclass[journal]{IEEEtran}

\usepackage{cite}
\usepackage{array}
\usepackage{arydshln}
\usepackage{comment}

\ifCLASSINFOpdf
  \usepackage[pdftex]{graphicx}
\else
\fi

\usepackage {xcolor}
\usepackage{amsmath,amsthm,amsfonts,amscd}  

\usepackage{algorithmic}
\usepackage{algorithm}

\usepackage{array}
\usepackage{multirow}

\ifCLASSOPTIONcompsoc
  \usepackage[caption=false,font=normalsize,labelfont=sf,textfont=sf]{subfig}
\else
  \usepackage[caption=false,font=footnotesize]{subfig}
\fi

\usepackage{url}

\begin{document}
%
\title{Fast Algorithm for Full-wave EM Scattering Analysis of Large-scale Chaff Cloud with Arbitrary Orientation, Spatial Distribution, and Length}
%
%
%

\author{Chung~Hyun~Lee, 
Dong-Kook Kang, 
Kyoung Il Kwon, 
Kyung-Tae Kim, and 
Dong-Yeop Na
\thanks{
\quad Chung~Hyun~Lee is with the Department of Electrical and Computer Engineering, Ohio State University, Columbus, OH 43210, United States, Kyoung Il Kwon is with Agency for Defense Development (Republic of Korea), Daejeon, South Korea, and Dong-Kook Kang, Kyung-Tae Kim, and Dong-Yeop Na are with the Department of Electrical Engineering, Pohang University of Science and Technology, Pohang, Gyeongsangbuk-do 37673, South Korea (e-mail: lee.4542@osu.edu; dkkang@postech.ac.kr; kikwon@add.re.kr; kkt@postech.ac.kr; dyna22@postech.ac.kr)}}

%
%

\markboth{Preprint}%
{Shell \MakeLowercase{\textit{et al.}}: Bare Demo of IEEEtran.cls for IEEE Journals}
%

\maketitle

\begin{abstract}
We propose a new fast algorithm optimized for full-wave electromagnetic (EM) scattering analysis of a large-scale cloud of chaffs with arbitrary orientation, spatial distribution, and length. 
By leveraging the unique EM scattering characteristics in chaff clouds, we introduce the {\it sparsification via neglecting far-field coupling} strategy, which makes an impedance matrix block-banded and sparse and thereby significantly accelerates thin-wire approximate method-of-moments solvers.
Our numerical studies demonstrate that the proposed algorithm can estimate the monostatic and bistatic radar cross section (RCS) of large-scale chaff clouds much faster and with greater memory efficiency than the conventional multilevel fast multipole method (see Fig. \ref{fig:complexity}), while retaining the high accuracy.
This algorithm is expected to be highly useful for RCS estimation of large-scale chaff clouds in practical scenarios, serving as a cost-effective ground-truth generator.
\end{abstract}


\IEEEpeerreviewmaketitle

\section{Introduction}
\label{sec:introduction}
\IEEEPARstart{C}{haffs} are radiofrequency (RF) countermeasures frequently used by military aircraft, ships, and vehicles to interfere with enemy radar. 
Chaff is typically shaped as metallic cylinders with a high length-to-diameter ratio, designed to be half a wavelength to maximize the monostatic radar cross section (RCS) at the corresponding frequency.
In practice, chaff of different lengths is commonly used to achieve high RCS responses over a broad frequency range.
When released, millions of chaff particles are influenced by gravity, air drag, and turbulence, causing significant changes in their spatial distribution and orientation. 
Therefore, accurate and fast RCS estimation for dynamically changing chaff clouds\textemdash characterized by arbitrary spatial distribution, orientation, and different length\textemdash crucial for optimizing defensive aids systems.

Extensive research has been conducted on electromagnetic (EM) scattering analysis of chaff clouds. 
Marcus proposed analytic formulas for estimating coherent and incoherent bistatic RCS values for sparse chaff clouds with random spatial distribution and orientation \cite{marcus2015bistatic}. 
He also proposed the equivalent conductor method (ECM) for dense chaff clouds, treating them as an isotropic conducting medium \cite{marcus2007electromagnetic}. 
Seo \textit{et al.} later extended this by developing a generalized ECM methodology that considers chaff clouds as anisotropic conducting media \cite{seo2010generalized}.
The vector radiative transfer (VRT) method has also been used to analyze the bistatic RCS of chaff clouds in complex environments \cite{zuo2021bistatic,zheng2023RCS}.

On the other hand, other researchers have employed full-wave EM numerical methods based on first principles, such as the method of moments (MoM) with the multilevel fast multipole method (MLFMM) \cite{wickliff1974average,perotoni2010electromagnetic,lee2022analysis} and the finite-difference time-domain (FDTD) method \cite{pandey2013modeling}, to investigate the bistatic RCS of chaff clouds.
However, the extremely long simulation times\textemdash taking several tens of hours for scenarios with 1,800 to 12,000 chaffs\textemdash have hindered their practical application for real-time RCS estimation. 
As a result, the use of full-wave EM numerical methods has been primarily restricted to providing ground-truth data for small-scale chaff clouds.

In practice, a bundle of chaff explodes and rapidly disperses into the air, occupying a large volume and exhibiting very low spatial density within a few seconds. 
The use of iterative MoM solvers with MLFMM acceleration may not be ideal for rapid RCS estimation of such sparse chaff clouds.
This is because radiation and receive functions often have a wide bandwidth, necessitating extremely high plane-wave sampling rates, which leads to longer computation times \cite{gibson2024method,chew2001fast}.\footnote{Instead, the MLFMM acceleration is more appropriate for the analysis of dense chaff clouds.}
Hence, alternative acceleration methods are needed for accurate and efficient RCS estimation of large-scale chaff clouds.

In this letter, we propose a new fast algorithm for full-wave EM scattering analysis of large-scale clouds of chaffs with arbitrary spatial distribution, orientation, and length.
On top of the thin-wire approximate electric-field integral-equation method-of-moments (TW-EFIE-MOM) solver,\footnote{The validity of using TW-EFIE-MOM solvers for RCS estimation of chaff clouds has been demonstrated through experiments \cite{yang2024research}.} We introduce a simple yet highly effective acceleration strategy, called {\it sparsification via neglecting far-field coupling} (S-NFC), which makes the impedance matrix block-banded and sparse, thereby significantly accelerating the TW-EFIE-MoM solver.
The proposed algorithm is expected to be an effective tool for rapid and accurate RCS estimation of large-scale chaff clouds in practical scenarios, serving as a cost-effective ground-truth generator.
To the best of our knowledge, no such full-wave fast algorithm currently exists.


Note that the time convention $e^{j\omega t}$ is adopted for this work.

\section{Formulation}\label{sec:matinidea}
\subsection{Thin-wire approximate electric-field integral-equation method-of-moments formulation}
Consider an either vertically (V) or horizontally (H) polarized plane wave $\mathbf{E}_{inc}(\mathbf{r})$ is incident on a chaff cloud consisting of $N_c$ number of chaffs, each with arbitrary orientation, length, and spatial distribution in free space. 
Employing the thin wire approximation \cite{gibson2024method} where each chaff is assumed to be a thin perfect electric conductor wire, one can write the induced electric current density on the surface of $p$\textsuperscript{th} chaff by:
\begin{flalign}
\mathbf{J}_p(\mathbf{r}) 
= 
I_p(\mathbf{r})\hat{\mathbf{t}}_p/(2\pi D_p),\quad \mathbf{r} \in \Omega_p
\end{flalign}
where $D_p$ is a chaff diameter, $I_p(\mathbf{r})$ is an electric current, $\hat{\mathbf{t}}_p$ is a unit tangential vector, and $\Omega_p$ denotes the chaff region.
Dividing each chaff region into $N_v + 1$ segments and expanding electric currents in terms of rooftop basis functions defined on vertices,\footnote{Such segmentation creates $N_v + 2$ number of vertices but the first and last vertices are to be excluded since currents at the ends of a wire should be zero. Hence, $N_v$ number of degrees of freedom exists per chaff.} one can represent the electric current density induced in a chaff cloud by:
\begin{flalign}
\mathbf{J}(\mathbf{r}') \approx 
\sum_{p=1}^{N_c}
\sum_{q=1}^{N_v}
j_{p,q}\Lambda_{p,q}(\mathbf{r}')\hat{\mathbf{t}}_p
=
\sum_{n=1}^{N_{dof}}
j_{n}\Lambda_{n}(\mathbf{r}')\hat{\mathbf{t}}_n.
\label{eqn:current_expression}
\end{flalign} 
where the rightmost expression is for a global unknown index $n(p,q)$, which incorporates the chaff index $p$ and vertex index $q$, and $N_{dof}=N_c \times N_v$ is the total number of unknowns.
Without loss of generality, $D_p$ and $N_v$ in this work is fixed at $20~[\mu\text{m}]$ and $19$, respectively.
Substituting \eqref{eqn:current_expression} into the redistributed EFIE \cite{gibson2024method} and applying the Galerkin testing, one can obtain the following MOM formulation:
\begin{flalign}
\overline{\mathbf{Z}}\cdot \mathbf{j} = \mathbf{v}
\label{TW_EFIE_MOM}
\end{flalign}
where current unknowns are stored in a column vector $\left[\mathbf{j}\right]_n=j_n$, and elements of the impedance matrix $\overline{\mathbf{Z}}=\overline{\mathbf{Z}}^{(1)}+\overline{\mathbf{Z}}^{(2)}$ and the force vector $\mathbf{v}$ can be calculated from which
\begin{flalign}
\left[\mathbf{v}\right]_{m}
&=
-\frac{j}{\omega\mu_0}
\int_{\sigma_m}
\Lambda_{m}(\mathbf{r})
\left[
\hat{\mathbf{t}}_{m}
\cdot
\mathbf{E}_{inc}(\mathbf{r})
\right]
d\mathbf{r},\\
\left[\overline{\mathbf{Z}}^{(1)}\right]_{m,n}
&=
\int_{\sigma_m}
\int_{\sigma_n}
f_{m,n}(\mathbf{r},\mathbf{r}')
d\mathbf{r}'d\mathbf{r},
\\
\left[\overline{\mathbf{Z}}^{(2)}\right]_{m,n}
&=
-\frac{1}{k^2}
\int_{\sigma_m}
\int_{\sigma_n}
h_{m,n}(\mathbf{r},\mathbf{r}')
d\mathbf{r}'d\mathbf{r},
\end{flalign}
for $m,n=1,2,\cdots,N_{dof}$.
Here, $\sigma_m$ is a support of $m$\textsuperscript{th} basis function, and
\begin{flalign}
f_{m,n}(\mathbf{r},\mathbf{r}') &=\left[\hat{\mathbf{t}}_{m}\cdot\hat{\mathbf{t}}_{n}\right]\Lambda_{m}(\mathbf{r})\Lambda_{n}(\mathbf{r}')G(\mathbf{r},\mathbf{r}'),\\
h_{m,n}(\mathbf{r},\mathbf{r}') &= \left[\hat{\mathbf{t}}_{m}\cdot\nabla\Lambda_{m}(\mathbf{r})\right]\left[\hat{\mathbf{t}}_{n}\cdot\nabla'\Lambda_{n}(\mathbf{r}')\right]G(\mathbf{r},\mathbf{r}'),
\end{flalign}
where $G(\mathbf{r},\mathbf{r}')=e^{-jk |\mathbf{r}-\mathbf{r}'|}/(4\pi |\mathbf{r}-\mathbf{r}'|)$ denotes the free space scalar Green's function.
When building the impedance matrix, the singularity extraction can be easily done in the analytic fashion while the other off-diagonal elements can be computed numerically by using the Gaussian quadrature\cite{gibson2024method}.

\begin{figure}
\centering
\includegraphics[width=3.3in]{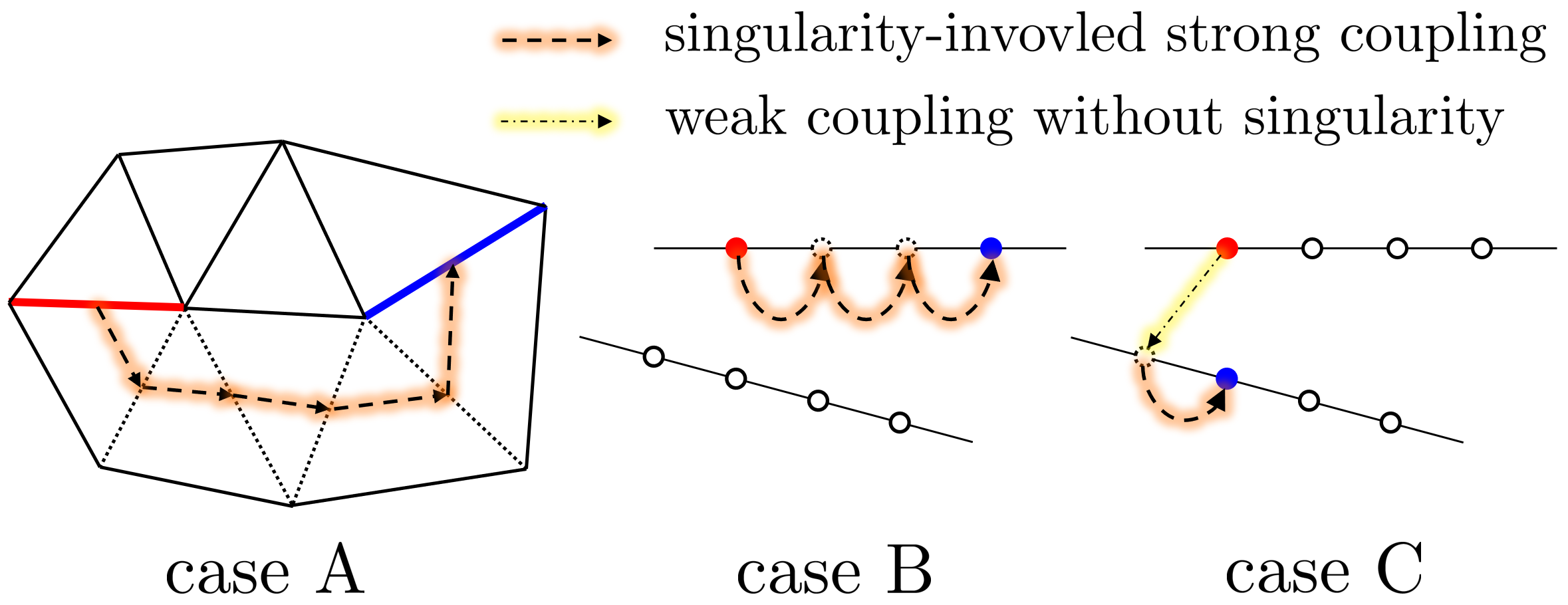}
\caption{Illustration of one of possible indirect information transmission routes between a pair of mesh currents for a 2-D mesh of a connected body, and an 1-D mesh two disconnected chaffs.}
\label{fig:interaction_map}
\end{figure}
\subsection{Sparsification via neglecting far-field coupling acceleration}
The EM scattering characteristics in a chaff cloud are strongly dependent on interactions between chaffs, influenced by the average pair distance $d_{mean} = (V_{cloud}/N_c)^{\frac{1}{3}}$, where $V_{cloud}$ denotes the chaff cloud volume. 
If $d_{mean} \geq 2\lambda$, mutual coupling effects can be ignored \cite{peebles1984bistatic,marcus2015bistatic}. We leverage this to reduce the computational complexity of the TW-EFIE-MOM solver by considering some important near-field coupling terms only, leading to a block-banded and sparse impedance matrix:
\begin{equation}
    \left[\overline{\mathbf{Z}}_{S-NFC}\right]_{m,n}
    =
    \begin{cases}
    \left[\overline{\mathbf{Z}}\right]_{m,n}, & d(\Omega_p,\Omega_{p'}) \leq d_c\\
    0, & \text{otherwise},
    \end{cases}
\label{equation:neardef}
\end{equation}
where $d(\Omega_p, \Omega_{p'})$ denotes the distance between centroids of the $p$\textsuperscript{th} and $p'$\textsuperscript{th} chaffs, and $d_c$ is the threshold distance for neglecting far-field coupling. 
Consequently, the S-NFC impedance matrix $\overline{\mathbf{Z}}_{S-NFC}$ contains self (diagonal) block matrices $\overline{\mathbf{Z}}_{p,p}$ (sized by $N_v \times N_v$) and mutual coupling (off-diagonal) block matrices $\overline{\mathbf{Z}}_{p,p'}$ (sized by $N_v \times N_v$) for $p \neq p'$. 
The average bandwidth $b_{mean}$ of $\overline{\mathbf{Z}}_{S-NFC}$ is related to $d_c$ and chaff cloud density, with the computational complexity of matrix-vector multiplication as $\mathcal{O}(b_{mean} N_{\text{dof}})$. The generalized conjugate residual (GCR) solver \cite{article:gcr1} with block-diagonal preconditioners is used, typically achieving convergence within a few tens of iterations.

It is important to note that using S-NFC may lead to incorrect results in typical surface integral equation scenarios involving a single target. 
A single target is usually assumed to have a closed surface, resulting in a closed mesh where all cells (vertices, edges, and faces) are connected. Consequently, all possible indirect interactions (or information transmission routes) between any pair of mesh currents include the strong coupling associated with the Green's function singularity, as illustrated in Fig. \ref{fig:interaction_map} (case A).
On the other hand, for chaff clouds, such strongly coupled indirect information transmission occurs only between pairs of vertices on the same chaff, not between vertices on different chaffs, as depicted in Fig. \ref{fig:interaction_map} (cases B and C). 
This distinction explains why the S-NFC acceleration is effective for RCS analysis of large-scale chaff clouds while maintaining high accuracy.

\section{Validation}
In this section, we consider several toy examples to validate the proposed algorithm, observing both deterministic and statistically averaged RCS behaviors.
\begin{figure}
\centering
\includegraphics[width=3.45in]{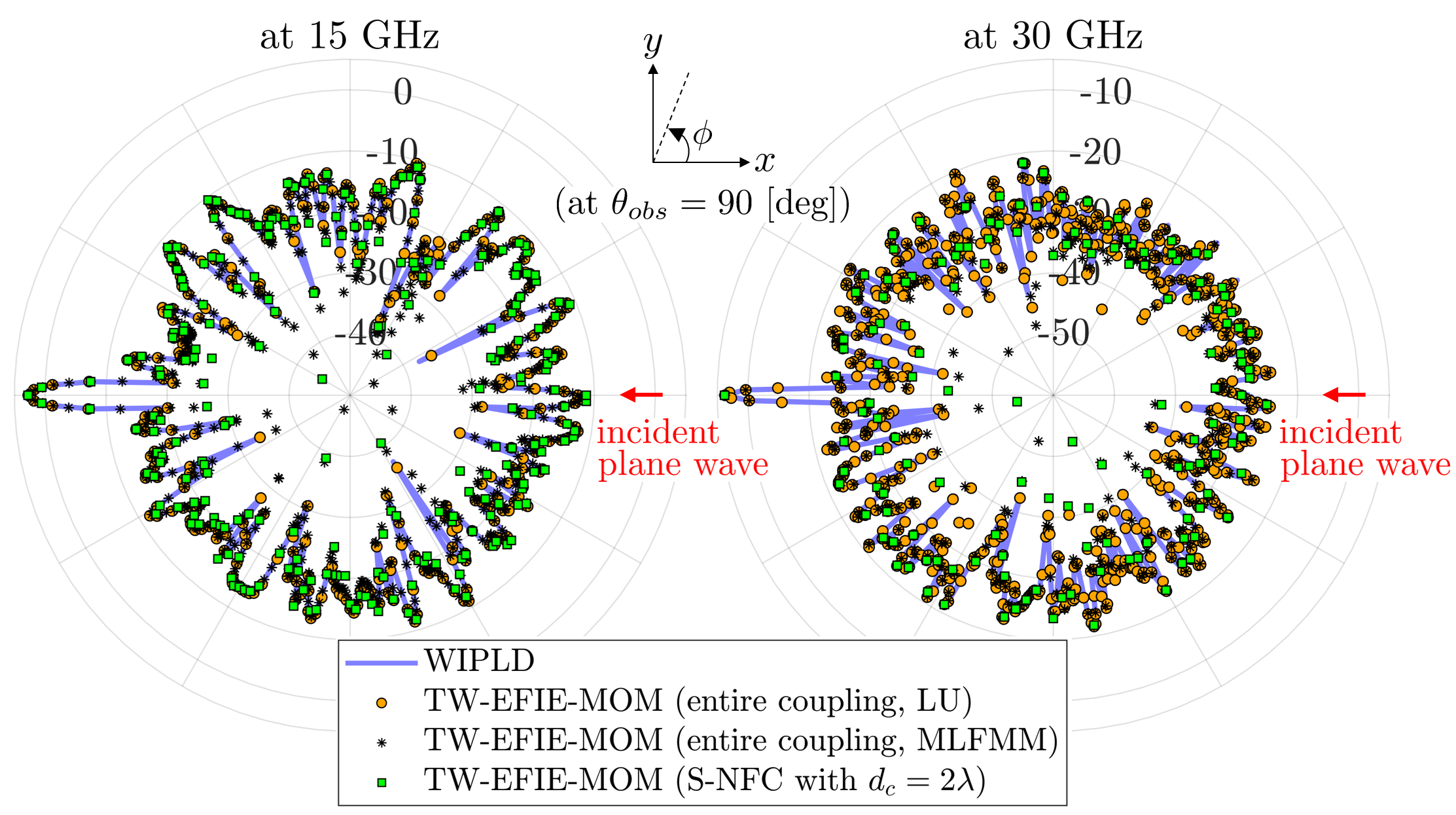}
\caption{Bistatic RCS $\sigma_{V,V}(\theta_{obs}=90^{o},\phi_{obs})$ [dBsm] for a V-polarized plane wave (coming from $x\rightarrow \infty$) incident on a chaff cloud consisting of $1,000$ identical chaffs ($10$ [mm] length) randomly distributed within a sphere with a radius of $0.2$ [m], each with a random orientation. 
}
\label{fig:WIPLD_validation}
\end{figure}
\label{sec:numerical-validation}
\subsection{Bistatic RCS for deterministic chaff cloud geometry}
We first compute bistatic RCS results for a fixed chaff cloud geometry and compare them with those obtained using the commercial software WIPL-D \cite{wipld2024}.
The chaff cloud consists of $1,000$ identical chaffs randomly distributed within a sphere with a radius of $0.2$ [m], each oriented randomly.
The length of each chaff element is set to $10$ [mm] (designed for the maximized monostatic RCS at $15$ [GHz]), and target frequencies are $15$ [GHz] and $30$ [GHz]; hence, $d_{mean}/\lambda_{15 \text{GHz}} \approx 1.61$.
Fig. \ref{fig:WIPLD_validation} shows the azimuthal sweep of the bistatic RCS on $xy$-plane (at $\theta_{obs} = 90^{o}$) when an incident plane wave is coming from $x\rightarrow \infty$ (i.e., $\theta_{inc}=90^{o}$ and $\phi_{inc}=90^{o}$).
It can be observed that the simulation results from WIPL-D and the proposed TW-EFIE-MOM methods across various implementations agree very well with each other.
Here, LU denotes a LU-decomposition direct solver.
As expected, the forward scattering ($\phi_{obs} = 90^{o}$) reaches its maximum, while the scattering intensity in all other directions is much smaller and maintains an almost uniform average with high fluctuations, due to the spherical shape of the chaff cloud.

\begin{figure}
\centering
\includegraphics[width=3.1in]{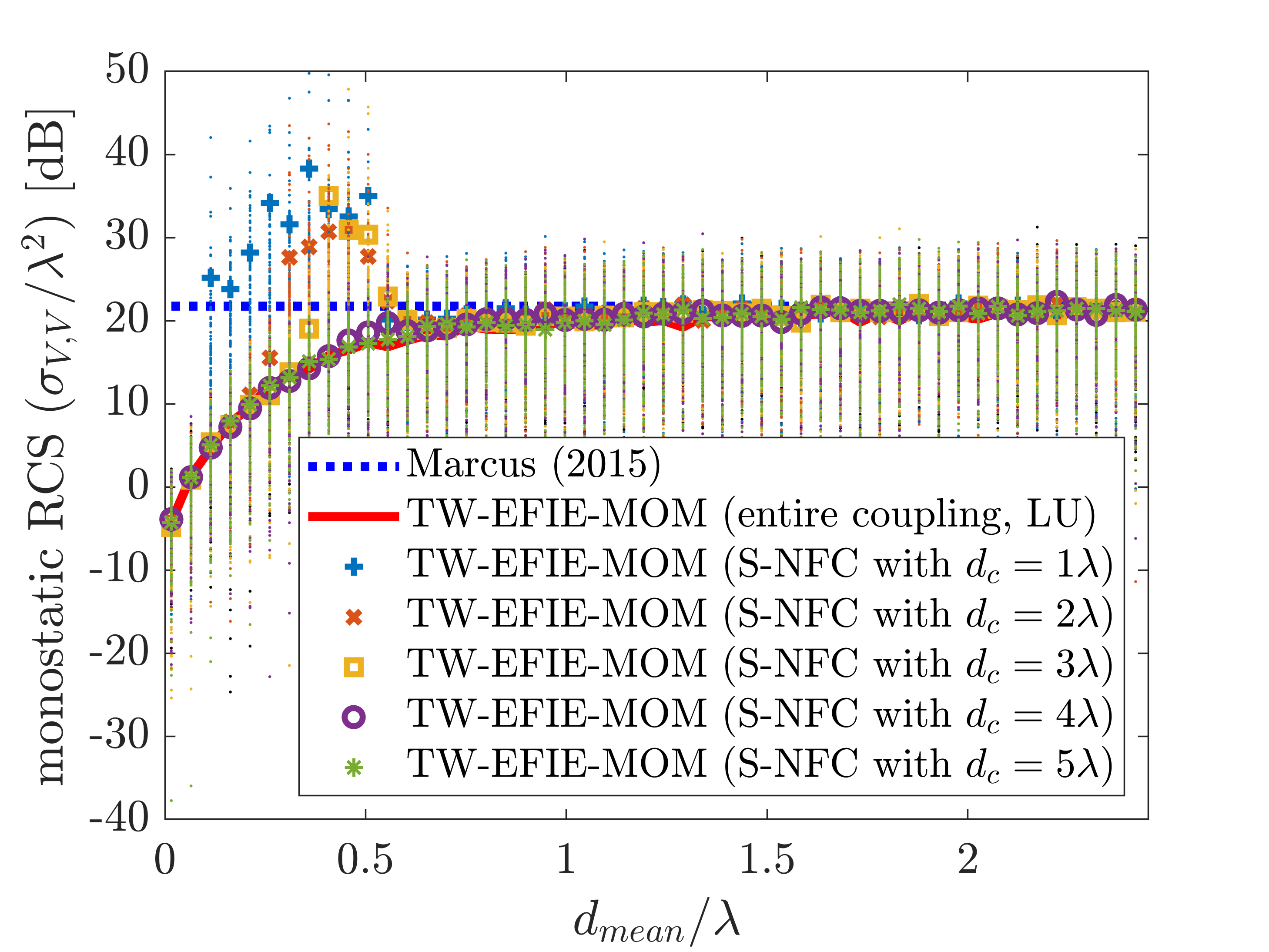}
\caption{Statistically averaged monostatic RCS $\sigma_{V,V}/\lambda^2$ [dB] versus different $d_{mean}$ from our full-wave methods (entire coupling (LU) and S-NFC with various $d_c$ values), compared with Marcus's prediction.}
\label{Convergence_marcus}
\end{figure}

\subsection{Statistically averaged monostatic RCS}
As shown in \cite{marcus2015bistatic}, for identical half-wavelength chaffs, random orientation, and uniform distribution with $d_{mean} \to \infty$, the coherent monostatic RCS converges to $0.15 N_c$. 
To verify this asymptotic behavior, we compute statistically averaged monostatic RCS values for various $d_{mean}$. For each $d_{mean}$, we conduct 100 independent trials with $1{,}000$ identical half-wavelength chaffs, randomly oriented and distributed. Fig. \ref{Convergence_marcus} shows the statistically averaged monostatic RCS vs. $d_{mean}$ from our full-wave methods (entire coupling (LU) and S-NFC with various $d_c$), compared with Marcus's prediction. 
As $d_{mean}$ increases ($d_{mean} \geq 0.5\lambda$), the RCS values converge to Marcus's prediction. 
For $d_c \geq 4\lambda$, no significant difference is observed between the entire coupling and S-NFC results, validating the S-NFC approach. 
For $d_{mean} < 0.5\lambda$, the monostatic RCS decreases nearly exponentially, consistent with \cite{marcus2007electromagnetic}. 
Note that TW-EFIE-MOM with S-NFC for $d_c \leq 3\lambda$ overestimates RCS due to locally overestimated Bragg resonance effects \cite{nandi2020controlling}. Increasing $d_c$ mitigates this, suggesting $d_c \geq 4\lambda$ for $d_{mean} \approx 0.5\lambda$, while $d_c = 2\lambda$ suffices otherwise.

\subsection{Frequency response}
\begin{figure}
\centering
\includegraphics[width=3.3in]{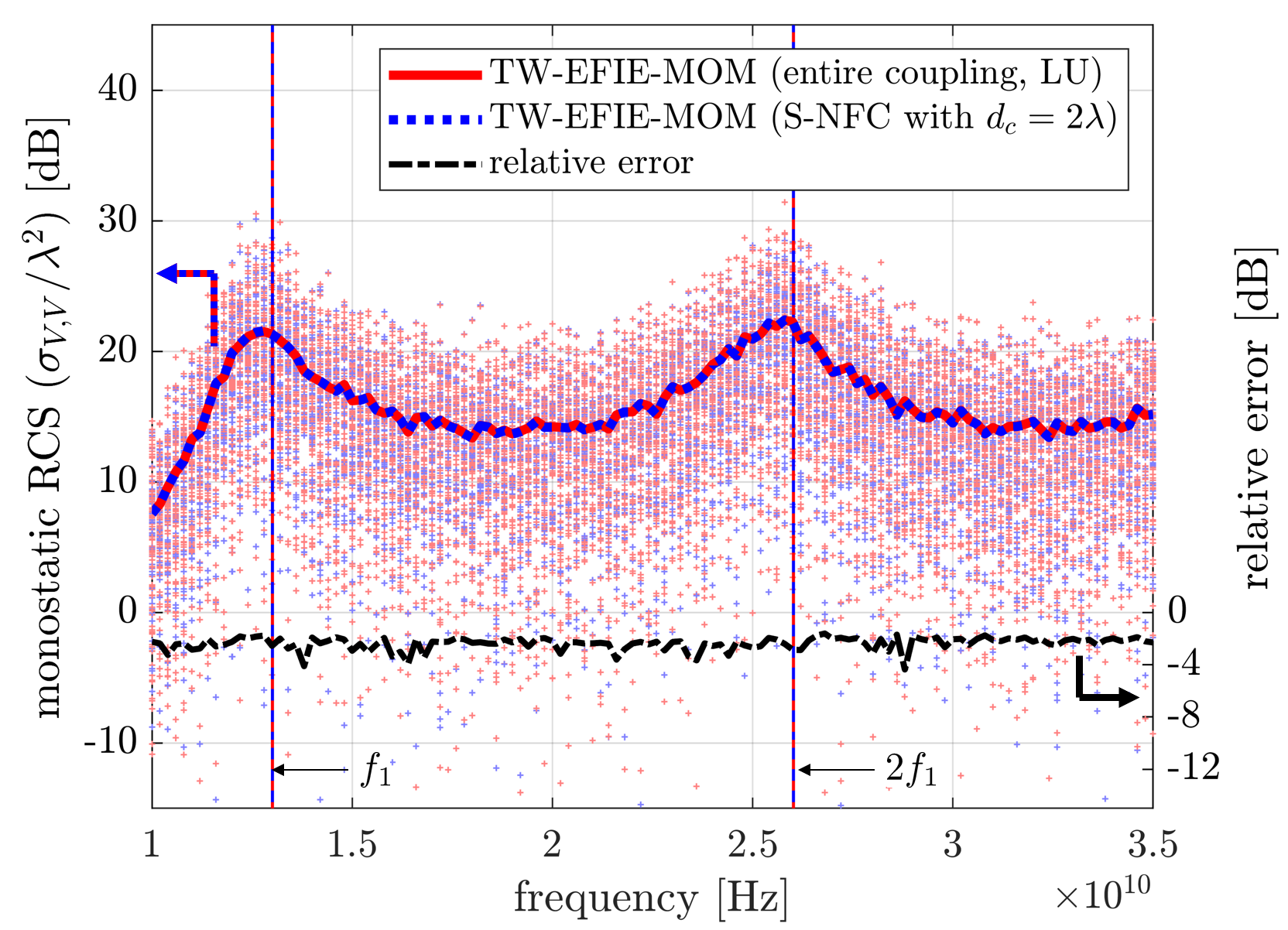}
\caption{Comparison of statistically averaged monostatic RCS $\sigma_{V,V}/\lambda^2$ [dB] for a chaff cloud consisting of $1,000$ identical chaffs over frequency obtained by using entire coupling (LU) and S-NFC with $d_c = 2\lambda$, and relative error.}
\label{fig:frequency_sweep1}
\end{figure}

We next consider the frequency response of the statistically averaged monostatic RCS $\sigma_{V,V}/\lambda^2$ [dB], and compare the results obtained using the entire coupling (LU) and S-NFC ($d_c = 2\lambda $) methods, as shown in Fig. \ref{fig:frequency_sweep1}. 
The chaff cloud consists of $1,000$ identical chaffs, each with a length of $l_1 = 0.01153$ [m], randomly orientated and distributed within a sphere of $a_c=10$ [m]. 
For each frequency, we again perform $100$ independent trials, using a new geometry per each trial. 
Two peaks are observed at $f_1 = c/(2l_1)$ and its double, corresponding to the resonant behaviors of the induced surface currents. 
The results from the entire coupling and S-NFC methods show excellent agreement, with the relative error, defined as $\log_{10}\left(\left|\sigma_{V,V}^{\text{(LU)}} - \sigma_{V,V}^{\text{(S-NFC)}}\right|/\left|\sigma_{V,V}^{\text{(LU)}}\right|\right)$, remaining below $-2$ [dB] across the entire frequency range.

\begin{figure}
\centering
\includegraphics[width=3.in]{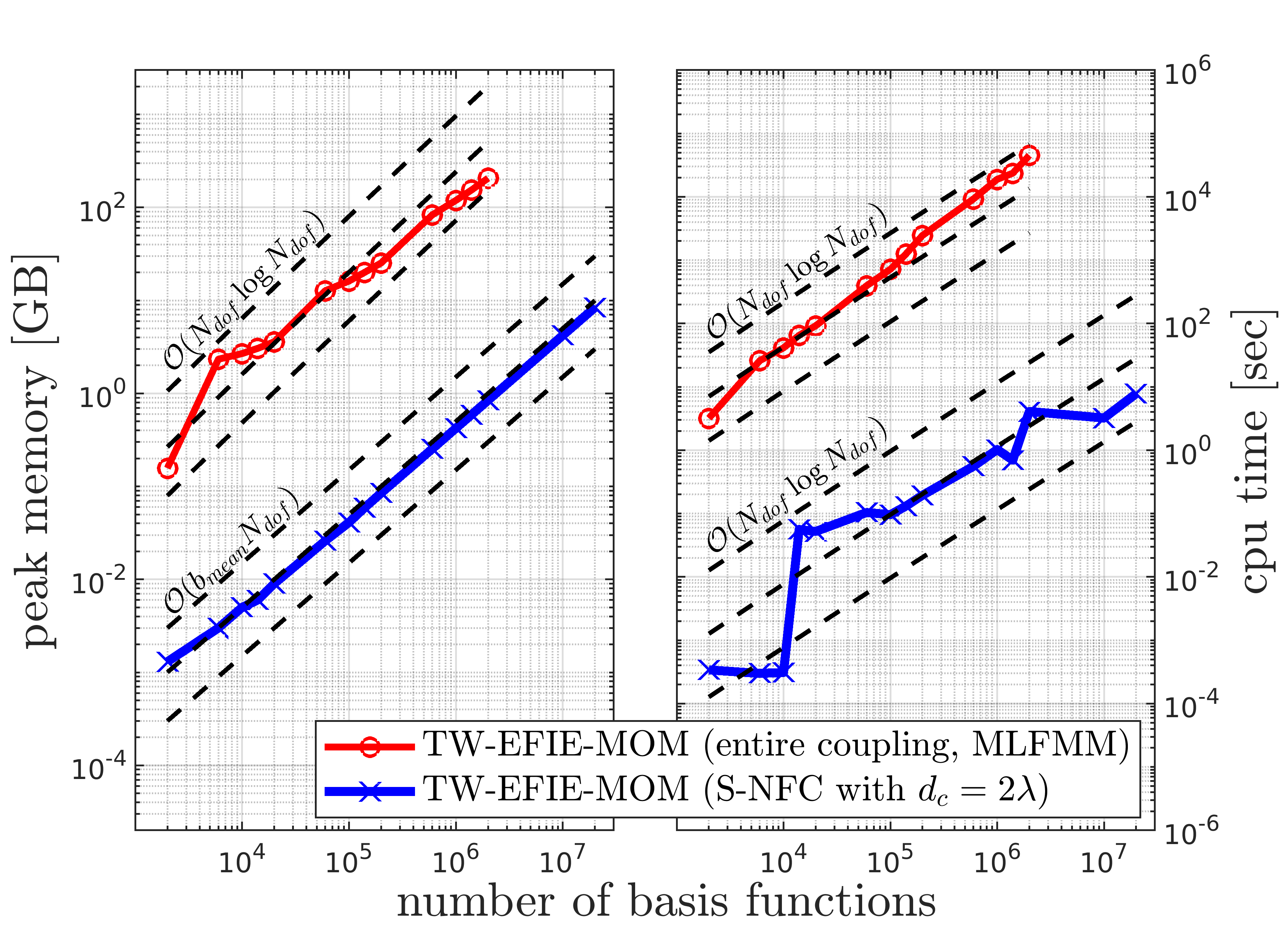}
\caption{Computational statistics: (left) peak memory during the entire solution process (right) CPU time per matrix-vector multiplication in iterative solvers.
}
\label{fig:complexity}
\end{figure}
We analyze the estimated complexity and speed-up of the proposed algorithm for randomly distributed chaff clouds within spherical domains. The chaff cloud radius ranges from $0.58$ to $12.6$ [m], with $N_c$ scaling from $10^2$ to $10^6$, maintaining a uniform density of $119.37$ chaffs/m$^3$. The frequency is fixed at $13$ [GHz], and data points are selected within the practical range for the number of basis functions. Experiments were conducted on a two-socket server with Intel Xeon Gold 6430 processors (64 cores, 1 [TB] memory).
Fig. \ref{fig:complexity} compares the CPU time and peak memory between the MLFMM and S-NFC implementations. 
As expected, both MLFMM show CPU time and peak memory complexity of approximately $\mathcal{O}(N_{def} \log N_{def})$. 
It appears to be less than $\mathcal{O}(N_{def} \log N_{def})$ for both CPU time and peak memory, although the very short CPU time of S-NFC introduces some instability, making it challenging to determine the exact complexity. 
With two million basis functions (i.e., $10^5$ chaffs), S-NFC also drastically cuts wall time for solving a single frequency point\textemdash $114$ [min] for MLFMM versus 12 [s] for S-NFC\textemdash achieving a substantial speed-up with a relative accuracy of 4.69\% in monostatic RCS. 
A critical factor in explaining performance difference is the excessively large number of plane-wave samples in MLFMM compared to the number of basis functions, especially in large-scale chaff cloud simulations.

\section{Numerical example}
\begin{figure}
\centering
\includegraphics[width=3.in]{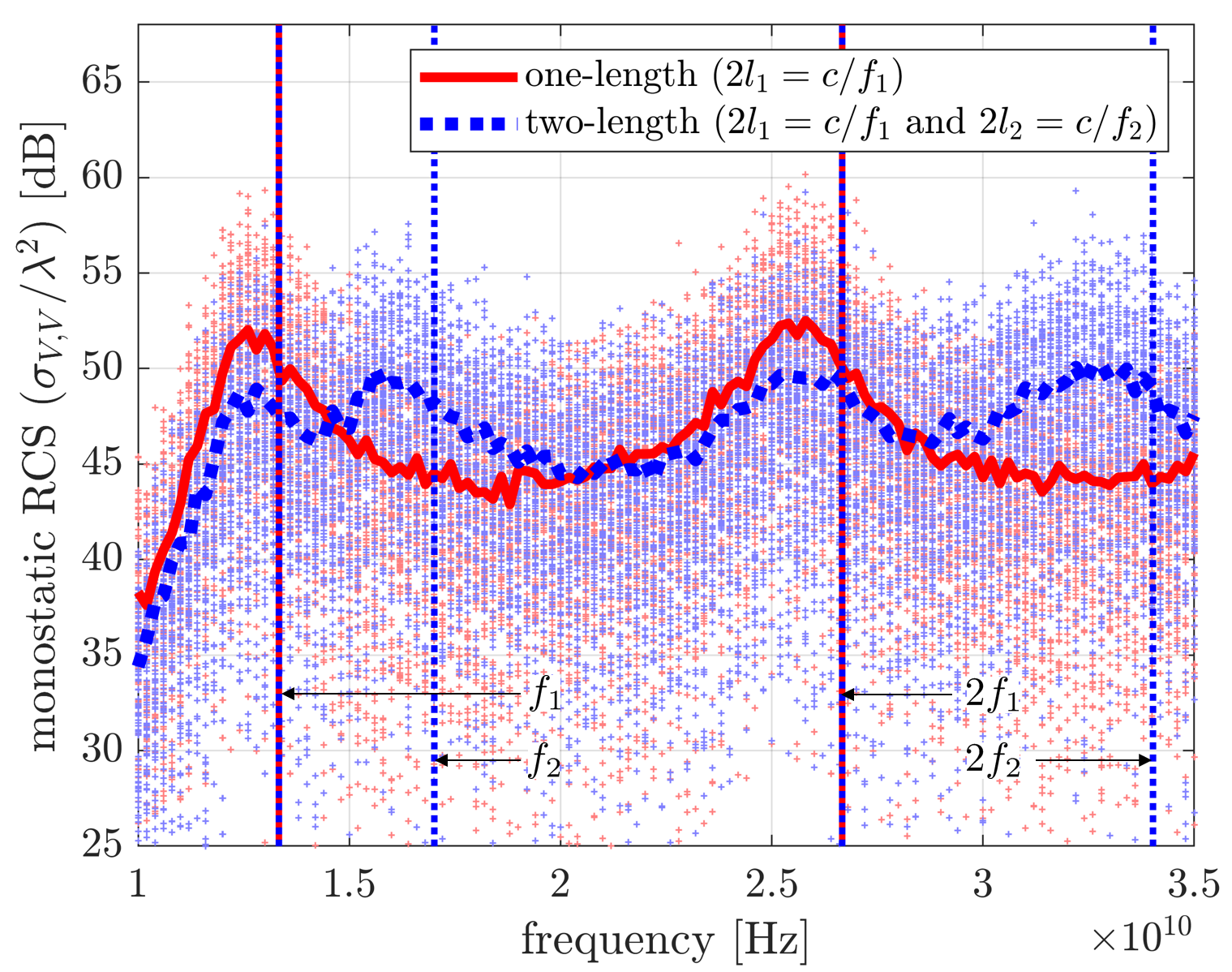}
\caption{Comparison of statistically averaged monostatic RCS $\sigma_{V,V}/\lambda^2$ [dB] of a chaff cloud composed of identical and non-identical $1$ million chaffs over frequency. Both results are obtained by using S-NFC with $d_c = 2\lambda$.}
\label{fig:frequency_sweep2}
\end{figure}
To validate the robustness of the proposed algorithm's speed-up, we consider a scenario with $10^6$ randomly oriented and distributed chaffs within a sphere of radius $a_c = 10$ [m]. Two cases are examined: (1) identical chaffs of length $l_1 = c/(2f_1) = 0.01125$ [m], and (2) non-identical chaffs, where half have length $l_1$ and the other half $l_2 = c/(2f_2) = 0.008817$ [m]. All experiments were conducted on the Nurion cluster (Intel Xeon Phi 7250, 68 cores) supported by Korea Institute of Science and Technology Information (KISTI). 
Fig. \ref{fig:frequency_sweep2} presents the statistically averaged monostatic RCS vs. frequency for both cases using the S-NFC method ($d_c = 2\lambda$). As expected, case (1) shows two peaks at $f_1$ and its harmonic, while case (2) exhibits four peaks at $f_1$, its second harmonic, $f_2$, and its second harmonic. Case (2) shows higher monostatic RCS across the frequency range, though peak values are slightly lower due to fewer chaffs contributing to resonance. This illustrates the importance of balancing broadband response and peak RCS when designing chaffs of different lengths. The proposed algorithm can handle chaffs of arbitrary lengths, though a two-length chaff cloud is used here to clearly illustrate peak separation.

\section{Conclusion and future works}\label{sec:conclusion}
We proposed a new acceleration technique, \textit{sparsification via neglecting far-field coupling} (S-NFC), integrated with a thin-wire electric-field integral-equation method-of-moments (TW-EFIE-MOM) solver. 
This method simplifies implementation by making the impedance matrix block-banded and sparse, enabling fast full-wave EM scattering analysis of large-scale chaff clouds with arbitrary orientations, spatial distributions, and lengths. The validity of S-NFC has been demonstrated through various examples, including the rapid computation of RCS for one million chaffs. This algorithm is highly useful for practical RCS estimation of large-scale chaff clouds, serving as a cost-effective ground-truth generator.

For high-density chaff clouds, S-NFC becomes equivalent to a direct solver, resulting in slower performance, where MLFMM is more efficient. Future work will focus on developing a hybrid acceleration method for the TW-EFIE-MOM solver. The approach will involve dividing large-scale chaff clouds into sections based on density, applying MLFMM to high-density sections ($d_{mean} \lambda \leq 1$) and S-NFC to low-density sections ($d_{mean} \lambda \geq 1$). This hybrid method aims to optimize matrix-vector multiplication at each iteration, with complexity bounded between $\mathcal{O}(b_{mean} N_{dof})$ and $\mathcal{O}(N_{dof} \log N_{dof})$.

\newpage

\ifCLASSOPTIONcaptionsoff
\newpage
\fi

\bibliographystyle{IEEEtran}
\bibliography{IEEEabrv,bibfile}

\end{document}